\documentclass[]{interact}

\usepackage{epstopdf}
\usepackage[caption=false]{subfig}

\usepackage[numbers,sort&compress]{natbib}
\bibpunct[, ]{[}{]}{,}{n}{,}{,}
\makeatletter
\def\NAT@def@citea{\def\@citea{\NAT@separator}}
\makeatother

\theoremstyle{plain}

\theoremstyle{definition}

\theoremstyle{remark}

\begin{document}

\articletype{Original Report}

\title{SrTiO$_3$ termination control: A method to tailor the oxygen exchange kinetics}

\author{
\name{Felix V.~E. Hensling \textsuperscript{a}\thanks{CONTACT F.~V.~E. Hensling. Email: f.hensling@fz-juelich.de}, Christoph Baeumer\textsuperscript{a}, Marc-André Rose\textsuperscript{b}, Felix Gunkel \textsuperscript{a,b}  and Regina Dittmann\textsuperscript{a}}
\affil{\textsuperscript{a}Peter Grünberg Institut 7 \& JARA-FIT, Forschungszentrum Jülich, 52425 Jülich, Germany; \textsuperscript{b}Institute of Electronic Materials (IWE2) \& JARA-FIT, RWTH Aachen University, 52074 Aachen, Germany}
}

\maketitle

\begin{abstract}
 We provide insights into the influence of surface termination on the oxygen vacancy incorporation for the perovskite model material SrTiO$_3$ during annealing in reducing gas environments. We present a novel approach to control to tailor the oxygen vacancy formation by controlling the termination. We prove that a SrO-termination can inhibit the incorporation of oxygen vacancies across the (100)-surface and apply this to control their incorporation during thin film growth. Utilizing the conducting interface between LaAlO$_3$ and SrTiO$_3$, we could tailor the oxygen-vacancy based conductivity contribution by the level of SrO termination at the interface. 
\end{abstract}


\section*{Impact Statement}
	
	For the first time the termination dependent oxygen exchange kinetics are reported for the perovskite model material SrTiO$_3$ and applied to the 2D electron gas model system LaAlO$_3$/SrTiO$_3$.
	
\section{Introduction}
Transition metal oxides have become a central topic in research over the past decade due to their manifold interesting properties.
\cite{C.N.R.RaoSolid1989,Christen2008} One of the most commonly used perovskites is SrTiO$_3$ (STO), which offers the advantage of a well known defect chemistry. Typical applications for STO as a functional material are resistive switching devices,\cite{Waser2009} catalysis\cite{Kawasaki2016} or thermoelectrics.\cite{Brooks2015}

Regardless of the application, the oxidation state of STO is a key parameter. In the field of thermoelectrics, oxygen vacancies are used to tailor the thermal conductivity of STO.\cite{Brooks2015,Breckenfeld2012a} When using STO for (photo-)catalysis, specific doping with oxygen vacancies is utilized to increase the activity.\cite{Tan2014,Mueller2015} Resistive switching of STO is based on the generation and redistribution of oxygen vacancies. Thus oxygen vacancies generated during growth or dedicated annealing steps define its switching properties.\cite{Janousch2007,Szot2006}  

The central role of oxygen vacancies for all STO applications has resulted in intense research efforts to understand and control their formation.\cite{Hensling2017,Hensling2018,Lee2016b,Sambri2012,Scullin2010,Chen2002b,Schneider2010,Gunkel2010,Moos1995b,Moos1997a,DeSouza2012,Xu2016a} Apart from the classic influence factors temperature and pressure\cite{Moos1995b,Moos1997a,DeSouza2012} it was found that e.g. UV radiation\cite{Hensling2018,Merkle2001,Merkle2002,Merkle2008,Leonhardt2002,Walch2017} plays a crucial role. STO substrates are easily reduced during thin film growth at low oxygen pressure. This is a drawback as it can mask the functional properties of the deposited thin film. Although a strong termination dependence of the oxygen exchange kinetics has been demonstrated for other perovskites,\cite{Tascon1981,Huang2014,Maiti2016} it has only been scarcely considered for STO(100),\cite{Alexandrov2009,DeSouza2012} which can exhibit a TiO$_2$-termination, a SrO-termination or a mixed termination.\cite{Kawasaki1993,Koster2000} The termination of STO especially plays a key role for applications, which rely on interfaces. Examples are the properties of magnetic heterojunctions,\cite{Zheng2010} interface dependent superconductivity,\cite{Rijnders2004a} and the formation of a two dimensional electron gas (2DEG), as observed at the interface of the model system LaAlO$_3$(LAO)/STO.\cite{Chen2011,Ohtomo2004,Breckenfeld2013,Herranz2007,Kalabukhov2007, Cancellieri2010,Huijben2009a,Nishimura2004,Nakagawa2006} The latter one is an outstanding candidate for an all-oxide field effect transistor. \cite{Woltmann2015,Goswami2015,Liu2012,Hosoda2013,Eerkes2013,Liu2015,Hurand2015}

Both properties, termination and oxidation state, thus play a central role in the field of STO applications. Yet their interplay has not been investigated so far, which is especially surprising as an influence of the termination layer on the oxygen exchange kinetics is well known for other perovskite systems.
In this work we will thoroughly investigate the role of the STO termination on the oxygen exchange kinetics. We find that the TiO$_2$-termination is more favorable to form oxygen vacancies, while the SrO-termination can completely suppress their formation. Hence, we present a new method to tailor the oxygen vacancy formation in SrTiO$_3$, namely by a precise control of the termination. In the course of this we will present evidence for an instability of the SrO termination under ambient conditions. Finally we will apply this knowledge to LAO/STO and show that we can tailor the resistivity of both, the crystalline system grown under reducing conditions and the amorphous system. As the resistivity for both system depends on oxygen vacancies, we thus prove termination control is a new method for controlling the formation of oxygen vacancies.

\section{Experimental}

To systematically control different degrees of SrO-termination, SrO was deposited on TiO$_2$-terminated STO (treatment with buffered HF)\cite{Kawasaki1993,Koster2000} from a SrO$_2$ target with a laser fluence of 0.81~$\frac{\textnormal{J}}{\textnormal{cm}^{2}}$ at an oxygen pressure of 10$^{-7}$~mbar and $800~^\circ\text{C}$ substrate temperature with a target substrate distance of 44~mm. The SrO coverage was controlled using reflective high energy electron diffraction. The deposition of both, crystalline and amorphous LAO, was performed at the same target substrate distance, at 10$^{-4}$~mbar and with a laser fluence of 1.3~$\frac{\textnormal{J}}{\textnormal{cm}^{2}}$ from a single crystalline LAO target. Amorphous LAO was grown at room temperature and crystalline LAO at $800~^\circ\text{C}$ with subsequent quenching (cool down time to below 400~$^{\circ}$C was about 40~s). The crystalline thin films were $8$ unit cells thick, the amorphous about 12 unit cells.

For the \textit{in situ} annealing process samples were annealed under oxidizing conditions after termination to minimize adsorbates. Subsequently the annealing experiment was performed for 1~h at an oxygen pressure of 10$^{-6}$~mbar at $800~^\circ\text{C}$ with subsequent quenching (cool down time to below 400~$^{\circ}$C was about 40~s).

The \textit{ex situ} samples were stored at room temperature under ambient conditions for 60~h before being exposed to the same annealing conditions. 

Electrical characterization at room temperature was performed using a \textit{Lakeshore 8400 Series} Hall measurement setup. Low temperature electrical characterization was performed with a physical property measurement setup. The XPS is a \textit{PHI 5000 Versa Probe} and the AFM a \textit{Omicron VT AFM XA}. The photoemission angle was $45^\circ$ and the spectra were fitted using \textit{Casa XPS} with a Shirley background and a convolution of Gaussian and Lorentzian line shape.

\section{Results}

\subsection{Interplay of termination and oxygen vacancy incorporation}

In order to investigate the termination dependent oxygen vacancy incorporation, STO single crystals were SrO- and TiO$_2$-terminated selectively. After termination, we applied the \textit{in situ} annealing process at 10$^{-6}$~mbar and $800~^\circ\text{C}$. The subsequent quenching of the samples preserves the defect equilibrium achieved at high temperature.  

\begin{figure}[h]
	\centering
	\includegraphics[width=\linewidth]{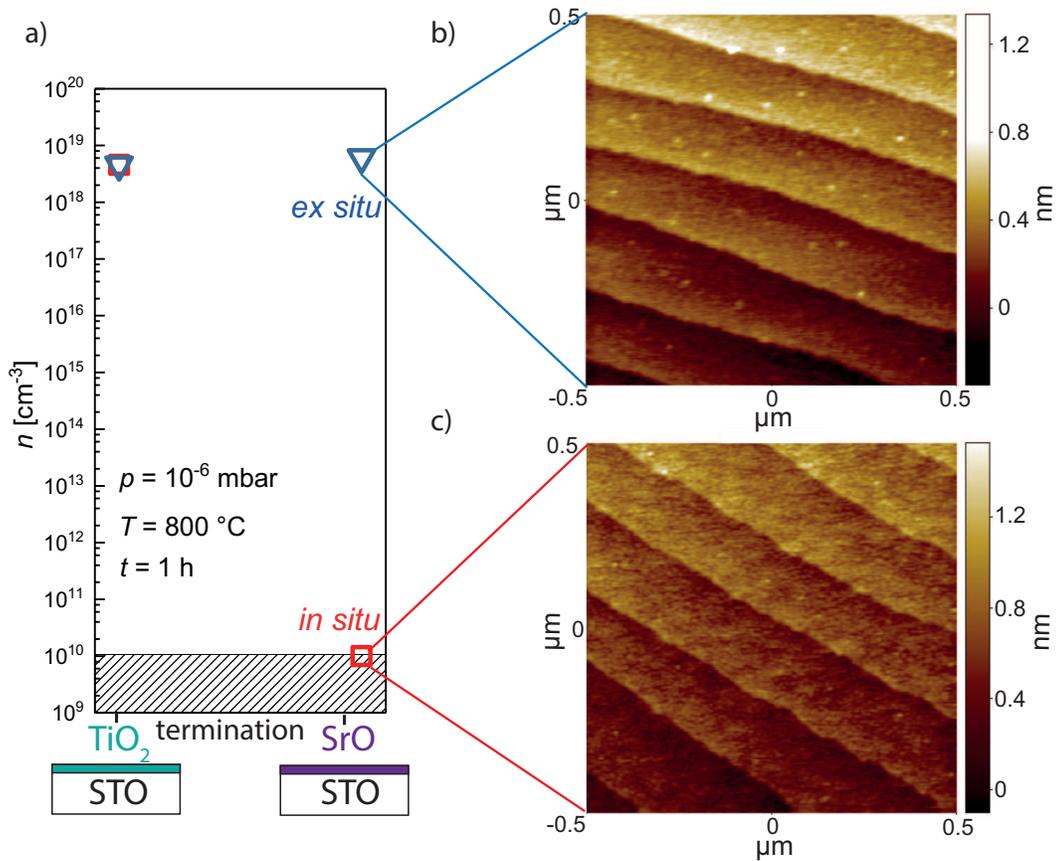}
	\caption{a) Carrier concentration of TiO$_2$ terminated (left) and SrO terminated (right) STO after 1~h of annealing at 10$^{-6}$~mbar oxygen pressure and $800~^\circ\text{C}$ for an \textit{in situ} sample (red) and a sample stored 60~h in air (blue). b) the topography of the SrO terminated sample after 60~h of air storage and c) the topography of a SrO terminated sample measured \textit{in situ} before storage in air. The topography in b) was measured for the sample presented in c), after that sample was exposed to air for 60~h.}
	\label{fig:STO}
\end{figure}

Figure \ref{fig:STO} a) shows the resulting carrier concentration $n$, which is a measure for the oxygen vacancy concentration, for TiO$_2$ and SrO terminated STO (red square). The resulting carrier concentration for the TiO$_2$ terminated sample is $5\times10^{18}$~cm$^{-3}$, a typical value observed for the applied  conditions.\cite{Hensling2017,Frederikse1964a} The carrier concentration of the SrO terminated sample after the same treatment is, however, below the measurement limit ($<10^{10}$~cm$^{-3}$). This is a first hint to a termination dependency of the oxygen vacancy formation.  

The observed behavior, however, changes drastically after SrO terminated STO was exposed to ambient conditions for 60~h. The annealing of \textit{ex situ} samples results in high carrier densities independent of the termination (blue triangles). Ambient storage thus affects the reduction of SrO-terminated STO, while the carrier concentration of TiO$_2$-terminated STO remains unchanged.  

Concomitant to the reduction behavior of SrO-terminated STO, also the topography changes, when exposed to ambient conditions . Figure \ref{fig:STO} c) shows the topography of an \textit{in situ} annealed sample and figure \ref{fig:STO} b) shows the topography of the same sample, but stored 60~h in ambient. While the vicinal surface is atomically flat in the beginning, we can observe features of about 0.2~nm height  decorating the unit cell step terraces after ambient storage. The topography of TiO$_2$-terminated STO showed no change after ambient storage.

\subsection{Stability of the SrO termination}

\begin{figure*}[h]
	\centering
	\includegraphics[width=\linewidth]{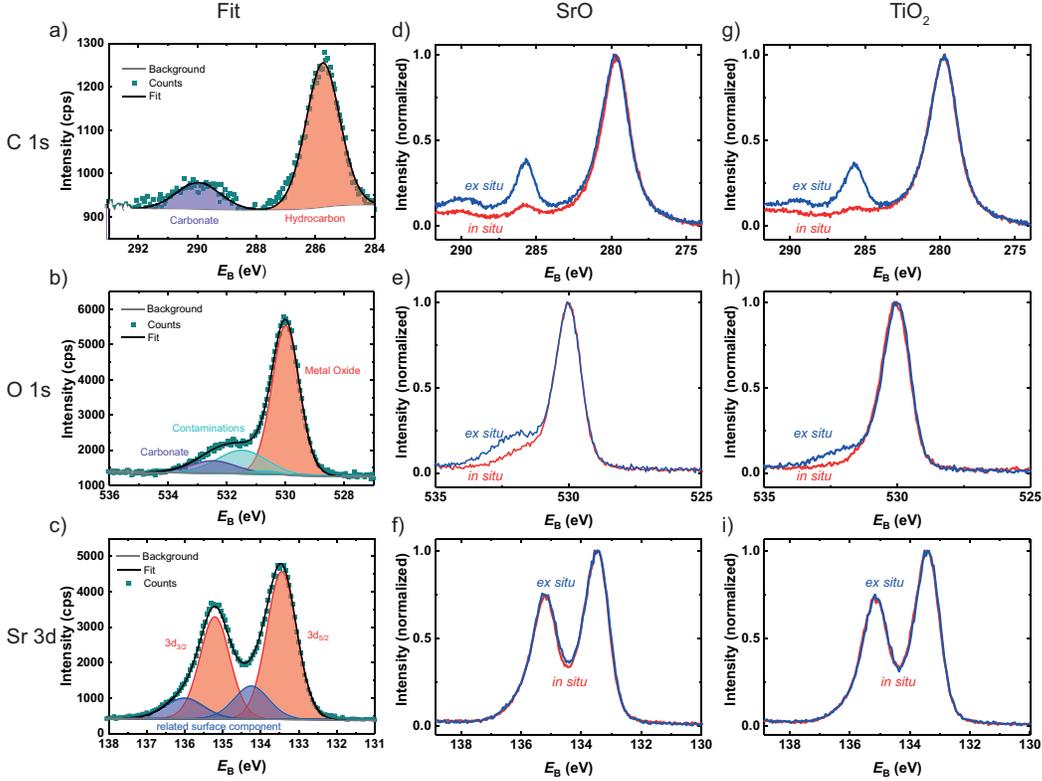}
	\caption{Left column shows the fits for a) the C 1s, b) O 1s and c) Sr 3d spectra. The center column shows the comparison of normalized spectra before (\textit{in situ}), blue, and after (\textit{ex situ}), red, ambient exposure for a SrO-terminated sample, the right column for a TiO$_2$-terminated sample, respectively. For the C 1s spectrum the change by ambient exposure is more significant for the d) SrO-terminated sample than the g) TiO$_2$-terminated sample. The same is observed for the SrO- e) and TiO$_2$-termination h) for the O 1s spectrum and the Sr 3d spectrum, f) and i), respectively.}
	\label{fig:AFMXPS}
\end{figure*}

We investigated the changes of the surface configuration during exposure to ambient conditions further using \textit{in situ} and \textit{ex situ} XPS in order to investigate the surface chemistry before and after ambient exposure. The C 1s (Figure \ref{fig:AFMXPS} top), O 1s (Figure \ref{fig:AFMXPS} center) and Sr 3d spectra (Figure \ref{fig:AFMXPS} bottom) were recorded. The C 1s spectrum is of interest as exposure to ambient is expected to give significant rise to adventitious carbon.\cite{Barr1995,Swift1982} The O 1s and Sr 3d spectra are of interest to probe chemical changes in the SrO termination layer. The center and right column show the comparison of the spectra before (\textit{in situ}) and after (\textit{ex situ}) exposure to the ambient for a SrO and a TiO$_2$-terminated sample, respectively.

The fits in the left column of Figure \ref{fig:AFMXPS} are representatively depicted for the SrO-terminated sample and were obtained in the same manner for the TiO$_2$-terminated sample. The chemical information is gained from these fits. The comparisons in the center and right column elucidate the differences occurring for both terminations before and after exposure to the ambient. The C 1s spectra (Figure \ref{fig:AFMXPS} a)) are fitted with a carbonate component for the highest binding energy ($E_B$) and a hydrocarbon component ($E_B\approx286~$eV). The O 1s spectra are fitted using 4 components (Figure \ref{fig:AFMXPS} b)). The lowest $E_B$ component represents metal oxide bonds in the STO bulk. The highest $E_B$ component represents carbonates.\cite{Shchukarev2004,Lam2017,Crumlin2012} Both of the intermediate $E_B$ can be ascribed to hydroxides and other non-carbonate contaminations, and are referred to as contaminations peaks. The Sr 3d spectra are composed of a doublet from the STO bulk and a second, surface-related doublet at higher binding energies, as is typically observed for STO (Figure \ref{fig:AFMXPS} c)).\cite{Szot2000} 

In order to estimate changes of the surface stoichiometry, we compared the the peak areas of the different core-levels. In case of the C 1s spectra we use the area ratio of the C 1s and Sr 3p$_{3/2}$ peaks (Figure \ref{fig:AFMXPS} d) and g)) to obtain a C/Sr ratio. We further use the area of the carbonate and the hydrocarbon peak to obtain a relative carbonate contribution. For the O 1s spectra, we use the areas of the carbonate and the contamination peaks in relation to the metal oxide peak to obtain a relative carbonate contribution and a non-carbonate contamination concentration, respectively. The relative contribution of the Sr related surface component to the Sr 3d signal is obtained from the ratio of the doublet at high binding energy and the doublet from the bulk STO.

\begin{table*} [h]
	\caption{\label{tab:XPS}Ratios obtained from XPS fits for \textit{in situ} and \textit{ex situ} samples, for both TiO$_2$- and SrO-termination, respectively.}
	\begin{tabular}{p{0.4\linewidth}||p{0.1\linewidth}|p{0.1\linewidth}||p{0.1\linewidth}|p{0.1\linewidth}}
		
		&\multicolumn{2}{c||}{\textit{in situ}}&\multicolumn{2}{|c}{\textit{ex situ}}\\ \hline \hline
		termination&TiO$_2$&SrO&TiO$_2$&SrO\\ \hline \hline
		C/Sr ratio (C 1s)&2.1~\%&8.6~\%&40~\%&40~\%\\ \hline
		Carbonate contribution (C 1s)&-&-&3.0~\%&6.5~\%\\ \hline
		Contamination concentration (O 1s)&24~\%&21~\%&28~\%&30~\%\\ \hline
		Carbonate contribution (O 1s)&-&-&7.5~\%&9.5~\%\\ \hline
		Sr related surface component (Sr 3d)&18~\%&18~\%&18~\%&21~\%\\
		
	\end{tabular}
\end{table*}

These ratios can be found in Table \ref{tab:XPS} for SrO and TiO$_2$-terminated samples, measured \textit{in situ} and after 60~h of ambient exposure (\textit{ex situ}). After air exposure, the C/Sr ratio increases markedly, to 40\%, for both, TiO$_2$- and SrO-termination. For both samples there is no discernible C 1s carbonate component, when measuring \textit{in situ}. Ambient exposure gives rise to this component, 3.0\% for the TiO$_2$-termination and 6.5\% for the SrO-termination. The carbonate component is thus markedly the highest for an ambient exposed SrO-termination. Similarly, the O 1s spectra do not exhibit a carbonate component when measuring \textit{in situ}, but after exposure to ambient, this component is 7.5\% for the TiO$_2$-termination and 9.5\% for the SrO termination. Again the carbonate component is thus the most pronounced for an ambient exposed SrO-termination. The 3d surface component of the Sr 3d spectra is 18\% for the TiO$_2$-termination \textit{in} and \textit{ex situ} and for the \textit{in situ} SrO-termination. Ambient exposure of the SrO-termination increases this spectral weight (21\%). This means that the 3d surface component of the Sr 3d spectra only changes after ambient exposing the SrO-terminated STO. 

The strong increase of the C/Sr ratio for both terminations is typical for XPS measurements after ambient exposure due to adventitious carbon.\cite{Barr1995,Swift1982,Piao2002,Miller2002} The more pronounced carbonate component of the C 1s and O 1s spectra for the SrO-termination points towards the formation of SrCO$_3$, naturally occurring when SrO reacts with CO$_2$ of the atmosphere.\cite{Ropp2013} This is also substantiated by the increase of surface component obtained from the Sr 3d spectrum after ambient exposure exclusively for the SrO-termination, as the component could be SrCO$_3$ related. It thus seems that the SrO-termination is instable under ambient conditions due to the formation of SrCO$_3$, eliminating the oxygen vacancy formation inhibiting effect.

\subsection{Application to LAO/STO heterostructures}

As we have developed a method to tailor the oxygen vacancy incorporation in STO by control of the the surface termination under UHV conditions, we next transferred this new method to a thin film system, namely LAO/STO. Since its discovery by  Ohtomo \textit{et al.}\cite{Ohtomo2004} LAO/STO is by far the most researched oxide 2DEG system.  

The formation of a 2DEG results from the evasion of the polar catastrophe by the transfer of half an electron to the TiO$_2$-terminated $n$-type interface, ultimately resulting in a 2DEG.\cite{Nakagawa2006} In the same way, one would expect half a hole to be transfered into the p-type interface for a SrO-termination. The potential build-up is, however, in this case compensated by positively charged oxygen vacancies rather than holes.\cite{Lee2018} This results in an insulating interface for the SrO-termination.\cite{Nakagawa2006} The $n$-type conducting interface was shown to prevail for interfaces with up to 83\% SrO-termination, which can be explained by the formation of the 2DEG in the TiO$_2$-terminated areas and percolation paths in between those areas.\cite{Huijben2009a,Nishimura2004} 

If LAO/STO structures are grown at low oxygen pressures the conduction mechanism changes. We have previously shown that the growth of crystalline LAO at oxygen pressures $\leq10^{-3}$~mbar results in a shift from 2DEG conductivity to bulk conductivity, when quenching the sample immediately after growth.\cite{Hensling2017,Xu2016a} The appearance of bulk conductivity in crystalline LAO/STO can be explained by the incorporation of oxygen vacancies in the STO bulk, which contribute electrons to the conduction band. During its low pressure growth, LAO sucks oxygen from the underlying STO substrate resulting in the formation of oxygen vacancies in the STO.\cite{Schneider2010} The shift towards bulk conductivity can easily be identified as it is accompanied by a shift towards lower resistivity. \cite{Amoruso2011,Amoruso2012a,Breckenfeld2013,Herranz2007,Hwang2004,Sambri2012,Ohtomo2004,Xu2016a,Cancellieri2010,Hensling2017,Kalabukhov2007,Huijben2009a}

Similar to the bulk conductivity of crystalline LAO/STO grown at low pressures, the 2DEG conductivity of the interface between amorphous LAO and STO single crystals relies on the incorporation of oxygen vacancies into the STO lattice. The main difference is their confinement to the interface in case of amorphous LAO/STO.\cite{Sambri2012,Trier2013,Chen2011} Amorphous LAO/STO therefore can be expected to be sensitive to the oxygen vacancy incorporation at the interface. We thus expect to see differences in the conductivity depending on the termination for both, crystalline LAO/STO grown in the bulk-conducting regime (i.e. grown at low pressures)  and for amorphous LAO/STO.

\begin{figure}[h]
	\centering
	\includegraphics[width=\linewidth]{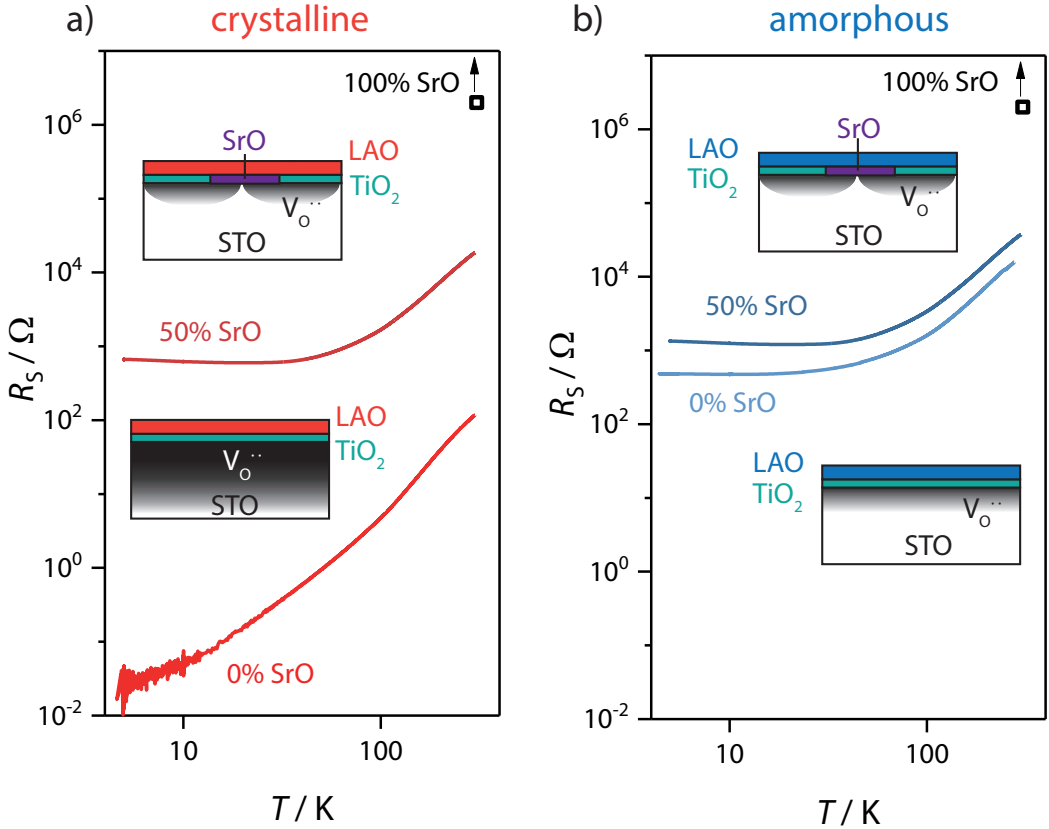}
	\caption{Sheet resistance in dependence of temperature of a) crystalline LAO/STO with different STO terminations and b) amorphous LAO/STO with different STO terminations. 100\% SrO-termination results in insulating behavior for both heterostructures. 50\% SrO-termination results in 2DEG conductivity for both, as does 0\% SrO termination in the amorphous case. The 0\% SrO-terminated crystalline LAO/STO shows metallic conductivity.}
	\label{fig:PPMS}
\end{figure}

Figure \ref{fig:PPMS} a) shows the sheet resistance of crystalline LAO/STO in dependence of temperature obtained for different STO terminations. The sheet resistance of LAO/STO with 0\% SrO-termination is about $100~\Omega$ at room temperature and about $10^{-2}~\Omega$ below 10~K. This corresponds to a dominant metallic bulk conduction of STO, as expected for these growth conditions.\cite{Xu2016a} An increase of the SrO-termination to 50\% results in an increase of the sheet resistance, to about $10^{4}~\Omega$ at room temperature and about $10^{3}~\Omega$ below 50~K. This is the typical temperature dependency of the sheet resistance for crystalline LAO/STO dominated by 2DEG conductivity.\cite{Amoruso2011,Amoruso2012a,Breckenfeld2013,Herranz2007,Hwang2004,
	Sambri2012,Ohtomo2004,Xu2016a,Cancellieri2010,Hensling2017,Kalabukhov2007,Huijben2009a} Increasing the SrO-termination further to 100~\% did, in agreement with observations reported in literature,\cite{Huijben2009a,Nishimura2004} result in insulating samples, whose sheet resistance is above the measurement limit (10$^8~\Omega$).   

Figure \ref{fig:PPMS} b) shows the sheet resistance of amorphous LAO/STO in dependence of temperature and STO termination. Amorphous LAO/STO with 0\% SrO-terminated STO has a sheet resistance of about $10^{4}~\Omega$ at room temperature and about $10^{3}~\Omega$ below 50~K. This is in good agreement with the temperature dependent resistivity of the 2DEG formed by amorphous LAO/STO, now solely relying on the incorporation of interface oxygen vacancies.\cite{Sambri2012,Trier2013,Chen2011} Increasing the SrO-termination of the STO substrate to 50\% results in a slight increase of the sheet resistance by a factor of two. A further increase to 100\% results, as for the crystalline case, in an insulating sample. Hence, only a negligible amount of oxygen vacancies has formed at the SrO-terminated interface. 

Figure \ref{fig:PPMS} shows that we can utilize the termination control of the STO substrate to tailor the sheet resistance in LAO/STO heterostructures. As the sheet resistance of both, crystalline LAO/STO grown in reducing conditions and amorphous LAO/STO, is defined by the oxygen vacancies\cite{Amoruso2011,Amoruso2012a,Breckenfeld2013,Herranz2007,Hwang2004,Sambri2012,Ohtomo2004,Xu2016a,Cancellieri2010,Hensling2017,Kalabukhov2007,Huijben2009a,Trier2013,Chen2011}, we have successfully tailored their incorporation. This also explains the shift from bulk dominated to interface dominated conductivity for crystalline LAO, when increasing the SrO-termination to 50\% (Figure \ref{fig:PPMS} a)). The oxygen vacancy incorporation is limited to the remaining TiO$_2$-terminated areas, as shown in the top sketch in Figure \ref{fig:PPMS} a). The resulting confined conductivity is comparable to the conductivity of amorphous LAO/STO (Figure \ref{fig:PPMS} b)), which is also dominated by oxygen vacancies confined to the interface, as shown in the according sketches. We can thus in the same way explain the increased resistivity for amorphous LAO/STO.

\section{Discussion}

Considering all the results described above we present a more complete picture of the role of the termination of STO for its oxygen exchange kinetics. Applying conditions known to be reducing for STO single crystals,\cite{Hensling2017,Frederikse1964a} we are only able to efficiently incorporate oxygen vacancies into the TiO$_2$-terminated single crystal. The SrO-terminated sample remains insulating. We conclude that the incorporation of oxygen vacancies is inhibited, as other kinetic and thermodynamic factors, e.g. the diffusion coefficient in the bulk, are not affected by termination.

\begin{figure}[h]
	\centering
	\includegraphics[width=\linewidth]{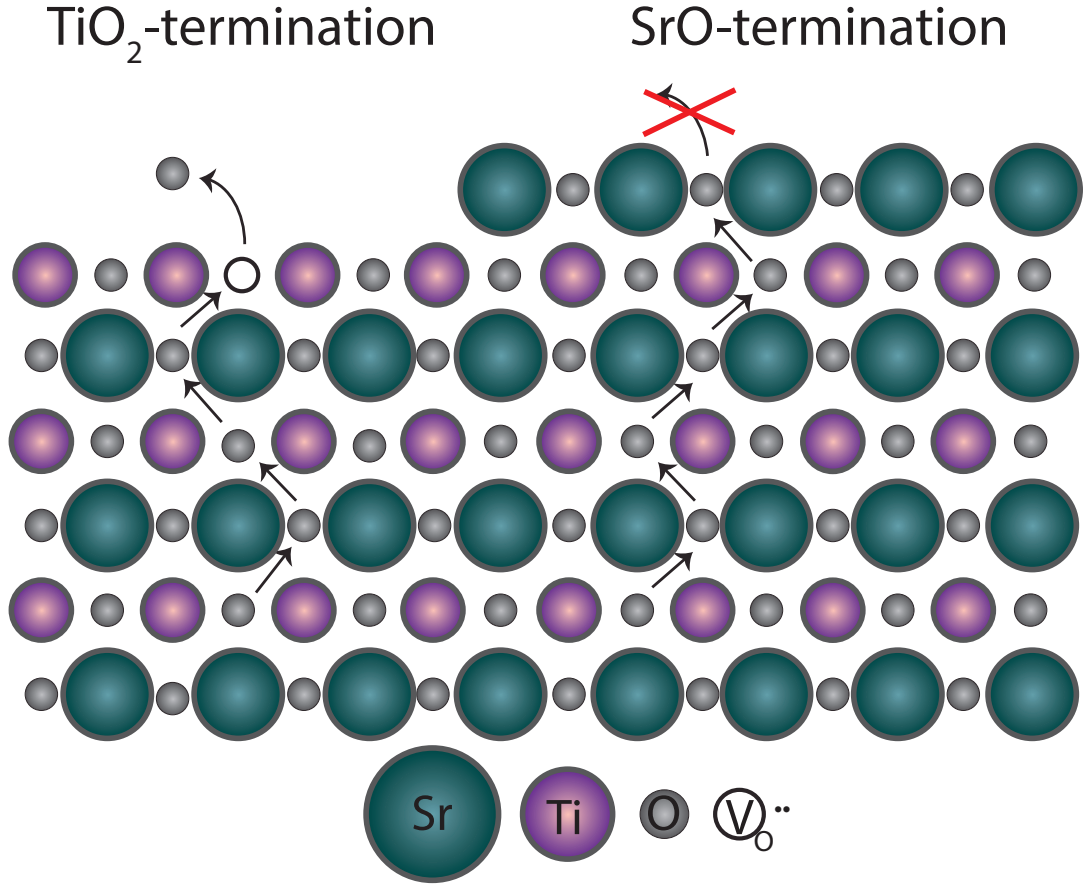}
	\caption{Schematic of the oxygen vacancy incorporation at the STO surface for different terminations. Due to the higher formation energy of oxygen vacancies for a SrO-termination, the oxygen vacancy incorporation is suppressed.}
	\label{fig:dis}
\end{figure}

The surface reaction of the oxygen vacancy incorporation is known to be a multi step process, which is influenced by several parameters.\cite{Merkle2002,Merkle2008} Considering that Alexandrov \textit{et al.}\cite{Alexandrov2009} have shown by \textit{ab initio} calculations that the formation energy of oxygen vacancies is lower for TiO$_2$-terminated STO as compared to SrO-terminated STO, it is conceivable that the termination has an influence on one or more steps of the surface reaction. 
\[
\textnormal{O}^x_\textnormal{O,surface} \rightleftharpoons \frac{1}{2}\textnormal{O}_2 + \textnormal{V}^{\bullet\bullet}_\textnormal{O} + 2\textnormal{e}^-
\]
In particular a high formation energy of oxygen vacancies in SrO-terminated STO could fully depress the oxygen vacancy incorporation at SrO-terminated surfaces. The schematic of this is shown in figure \ref{fig:dis}. 

Another possible mechanism behind the inhibited oxygen vacancy incorporation is SrO acting as a diffusion barrier for oxygen.\cite{Baeumer2015} However an increased diffusion barrier was only shown for SrO in a rock salt structure. For SrO in the perovskite structure of STO, the effect was not observed.\cite{Hensling2018a} We thus rule out the explanation of a SrO diffusion barrier, leaving the increased formation energy as decisive parameter.

The observed behavior, however, changed drastically after SrO-terminated STO was exposed to ambient conditions. Similar vacuum annealing of \textit{ex situ} samples now resulted in high carrier densities independent of termination, indicating that the blocking effect of the SrO termination layer was eliminated by air exposure. This was accompanied by the formation of morphological features, indicating a clustering of SrO related particles.\cite{Hensling2018a} Utilizing XPS measurements these particles were identified as SrCO$_3$.

If a SrO-termination inhibits the formation of oxygen vacancies, the formation of SrCO$_3$ can explain the elimination of the effect under ambient conditions. The formation of islands after ambient exposure results in pores in the oxygen blocking SrO-termination layer, revealing the TiO$_2$-terminated STO, which then allows pathways for the incorporation of oxygen vacancies. Moreover, the resulting SrCO$_3$ clusters do not necessarily have a high formation energy for oxygen vacancies. 

To profit from our new method to tailor the oxygen vacancy incorporation properties we applied it to the model application of the 2DEG at the LAO/STO heterointerface. Crystalline LAO/STO dominated by bulk conductivity and amorphous LAO/STO both rely on oxygen vacancy incorporation. In case of crystalline LAO/STO fabricated under reducing conditions we observe a transition from bulk conducting STO (0~\%) to interface conductivity (50~\%) and finally to insulating (100~\%) with increasing amount of SrO-termination . For amorphous LAO/STO we do as well observe a transition to insulating behavior, when increasing the SrO-termination to 100\%. This confirms the inhibition of the incorporation of oxygen vacancies we observed for STO single crystals. 

Considering crystalline LAO/STO this is especially interesting, as it was found that the conductivity in the initial report of 2DEG conductivity by Ohtomo \textit{et al.}\cite{Ohtomo2004} was in fact dominated by bulk conductivity, hence by oxygen vacancies.\cite{Breckenfeld2013,Herranz2007} Nevertheless, conductivity was only observed for TiO$_2$-terminated STO.\cite{Ohtomo2004} This effect could not be explained within the  polar catastrophe scenario. With our findings we are able to explain this phenomenon \textit{via} the inhibition of the oxygen vacancy incorporation for the SrO-terminated sample.

\section{Conclusions}
With this work we have provided a novel way to tailor the oxygen vacancy incorporation in STO by controlling its termination. We have demonstrated that a SrO-termination of STO completely inhibits the incorporation of oxygen vacancies for otherwise reducing conditions. By systematically controlling the termination of STO it is thus possible to tailor the areas of oxygen vacancy incorporation. Due to the widespread use of STO as a substrate this result is highly interesting. A specific application for which we employed this new method are LAO/STO 2DEG heterostructures. By doing so, we were not only able to directly influence the conductivity of these heterostructures, but did also improve their understanding. Further applications include, but are not limited to: i) oxides that require low oxygen pressure growth due to thermodynamic reasons (e.g. LaVO$_3$\cite{Hotta2007,He2012,Vrejoiu2016a}, EuTiO$_3$\cite{Shkabko2013a}), for which film properties would otherwise be masked by oxygen vacancies induced in STO; ii) metals, which induce the incorporation of oxygen vacancies at the interface.\cite{Santander-Syro2011}

\section*{Disclosure statement}
There are no conflicts to declare.

\section*{Acknowledgements}
We acknowledge funding from the W2/W3 program of the Helmholtz association. The research has furthermore been supported
by the Deutsche Forschungsgemeinschaft (SFB 917 ‘Nanoswitches’). F.G. and M.R. thank the DFG GU/1604. CB has received funding from the European Union’s Horizon 2020 research and innovation programme under the Marie Sklodowska-Curie grant agreement No 796142. We further thank R.A. de Souza and M. Müller for the helpful discussions.

\bibliographystyle{tfnlm}
\bibliography{library}

\end{document}